\documentclass[10pt]{article}
\begin{document}

\newcommand{\m}[1]{\ensuremath\mbox{\boldmath $#1$}}
\newcommand{\be}{\begin{equation}} \newcommand{\ee}{\end{equation}}
\newcommand{\ba}{\begin{eqnarray}} \newcommand{\ea}{\end{eqnarray}}
\newcommand{\nn}{\nonumber} \renewcommand{\bf}{\textbf}
\newcommand{\ra}{\rightarrow} \renewcommand{\c}{\cdot}
\renewcommand{\d}{\mathrm{d}} \newcommand{\diag}{\mathrm{diag}}
\renewcommand{\dim}{\mathrm{dim}} \newcommand{\D}{\mathrm{D}}
\newcommand{\integer}{\mathrm{integer}}\newcommand{\LL}{\lambda}
\newcommand{\R}{\mathbf{R}} \renewcommand{\t}{\mathrm{t}}
\newcommand{\T}{\mathbf{T}} \newcommand{\V}{\mathbf{V}}
\newcommand{\tr}{\mathrm{tr}} \newcommand{\cA}{\cal A}
\newcommand{\cB}{\cal B} \newcommand{\cC}{\cal C}
\newcommand{\cD}{\mathrm{\cal D}} \newcommand{\cF}{\cal F}
\newcommand{\cG}{\cal G} \newcommand{\cL}{\cal L}
\newcommand{\cO}{\cal O} \newcommand{\cT}{\cal T}
\newcommand{\cU}{\cal U} \newcommand{\s}{\,\,\,}
\renewcommand{\a}{\alpha} \renewcommand{\b}{\beta}
\newcommand{\e}{\mathrm{e}} \newcommand{\eps}{\epsilon}
\newcommand{\f}{\phi} \newcommand{\fr}{\frac} \newcommand{\g}{\gamma}
\newcommand{\h}{\hat} \renewcommand{\i}{\mathrm{i}}
\newcommand{\p}{\partial} \newcommand{\w}{\wedge} \newcommand{\x}{\xi}
\newcommand{\EE}{\vec E} \newcommand{\DD}{\Delta}

\input{epsf}


%
\title{Femto-Photography of Protons to Nuclei with Deeply Virtual
Compton Scattering }
\author{John P. Ralston$^{1,2}$ and Bernard Pire$^1$}

\maketitle
\begin{abstract}
{Developments in deeply virtual Compton scattering allow the direct
measurements of scattering amplitudes for exchange of a highly virtual
photon with fine spatial resolution.  Real-space images of the target
can be obtained from this information.  Spatial resolution is
determined by the momentum transfer rather than the wavelength of the
detected photon.  Quantum photographs of the proton, nuclei, and other
elementary particles with resolution on the scale of a fraction of a
femtometer is feasible with existing experimental technology.  }
\end{abstract}

 More than 40 years ago, elastic scattering of
relativistic electrons from protons by Hofstadter\cite{Hofs} {\it et
al } probed the dimensions of protons and nuclei.  On general
principles the scattering is governed by ``form factors", which
parameterize the difference between point-like scattering and the
observations.  In static non-relativistic approximations of the era, the form
factors were interpreted as ``\ldots determining the distribution of
charge and magnetic moment in the nuclei of atoms and of the nucleons
themselves''\cite{Friedman}, with the experiments receiving the Nobel
Prize in 1961.  The charge radius was found to be about 0.7
femtometer.  The neutron's form factor was later interpreted in terms
of a positively charged core surrounded by a negatively charged outer
shell.  In retrospect, these classic interpretations are open to
doubt.  Neither the impulse approximation nor the interpretation as
charge density applies so simply to hadronic physics in the regime of
the experiments.  Point-like structure now attributed to quarks has
been deduced indirectly, in conjunction with the development of
Quantum Chromodynamics and deeply {\it inelastic} scattering experiments (DIS).
The structure of hadrons as complex aggregates of quarks remains
rather mysterious.

Deeply virtual Compton scattering (DVCS)\cite{GPD} combines
features of the inelastic processes with those of an elastic process.
A relativistic charged lepton (electron, positron, or possibly a muon) is
scattered from a target nucleon or nucleus.  A real photon of
4-momentum $q_{\mu}'=(q_{0}', \vec q')$ is also observed in the final
state.  With $e(k),\, e'(k')$ denoting the initial and final electrons of
momenta $k, \, k'$ respectively, and $P, \, P'$ denoting the momentum of the
target, the process is $$e(k)+ P \ra e'(k')+P' +\gamma(q').$$ The net
momentum transfer $\Delta$ to the target is obtained by momentum
conservation, $\Delta ^{\mu} = k ^{\mu} -k^{` \, \mu } -q^{` \, \mu} $.
The real photon may be emitted by the lepton beam, in which case a
virtual photon of momentum $Q_{BH}^{\mu}=\Delta^{\mu}$ strikes the
target.  Otherwise the target emits the real photon and a virtual photon of
$Q_{VCS}^{\mu}=P'^{\mu}-P^{\mu}+q^{` \, \mu }$ strikes the target.  It
is straightforward to select events where all components of $\Delta
^{\mu}$ are small compared to $\sqrt{Q^{2}}$, with
$Q^{2}=-Q_{VCS}^{\mu}Q_{VCS, \, \mu} > GeV^{2}$.  These conditions
have recently been realized in experiments at HERMES\cite{HermesCEBAF}
at DESY and CEBAF\cite{HermesCEBAF} at JLab.

The physical interpretation is that the target is resolved by the
virtual photon on a spatial scale small compared to the target size.
A photon of high virtuality $Q^{2}$ selects a short-distance region of
the target: the spatial resolution is of order $\Delta b_{T} \sim
\hbar/ \sqrt{ Q^{2}}$.  In contrast, high energy real photons ($(q')
^2 =0$) are not well localized, and do not interact in a consistent
point-like manner described by simple impulse approximations.
Perturbative QCD (pQCD) can be applied to DVCS at large $Q^{2}$ ,
exploiting the short-distance resolution of the virtual photon,
despite the presence of a real photon in the reaction.
\cite{GPD,Ji97Radyushkin97}

Meanwhile the net momentum transfer $\Delta^{\mu}$ is independent, and
Fourier-conjugate to the {\it spatial location where the virtual
photon scattering event occurs.} Let us review this\cite{buniy}.  We
use a conventional Lorentz frame with coordinates (+, T, -) where
$P^{2}= 2 P^{+}P^{-}-\vec P_{T}^{2}$: \ba P^{\mu}=( P^{+},\, -\vec
\Delta_{T}/2,\, \frac{ m^{2}+ \Delta_{T}^{2}/4}{2P^{+}}); \nn \\
P'^{\mu}=( P^{+}+\DD^{+},\, +\vec \Delta_{T}/2,\, \frac{ m^{2}+
\Delta_{T}^{2}/4}{ 2P^{+}+2\DD^{+} }); \nn \\q^{\mu}=( Q_{VCS} ,\, 0,\,
Q_{VCS} )/\sqrt{2}; \nn \\
q'^{\mu}=( Q_{VCS}/\sqrt{2}-\DD^{+} , -\vec \DD_{T}, \frac{
\DD_{T}^{2}}{ 2(Q_{VCS}/\sqrt{2} -\DD^{+} )}).  \nn \ea We also write
$\DD^{\mu}= \zeta P^{\mu} + \Delta_{T}^{\mu}, $
$\DD^{\mu}\DD_{\mu}=t<0$.  Here $m$ is the target mass and $\zeta/2$
is denoted the `` skewness''.

The impulse approximation applies in the infinite momentum frames
$P^{+} \ra \infty$, $x_{Bj} =Q_{VCS}^{2}/2 P \cdot q$ fixed and not
too small.  The handbag diagram represents dominance by the {\it
quark-proton scattering amplitude} $\Phi$ upon which both diagonal and
off-diagonal (generalized) parton distributions are based.  Our
notation is \ba && \Phi^{\rho \sigma}(\kappa, \kappa')_{P,P'}= \nn \\
&&\int dx_{1} dx_{2} \, e^{i\kappa \cdot x_{1}-i\kappa' \cdot
x_{2}}<P',s'| T \psi_{\rho}(x_{2}) \bar \psi_{\sigma}(x_{1})|P,s>. 
\label{PhiDefined}\ea This object is relevant to the analysis in the
region\cite{DGPR} $-t_{min}<-t<-t_{max}\sim GeV^{2}$.  By momentum
conservation $\kappa'-\kappa=P'-P=\Delta$.  The quark fields are
evaluated at spatial coordinates $x_{1}, \, x_{2}$ and have Dirac
indices $\rho, \sigma$.  We may rewrite $$ e^{i\kappa \cdot
x_{1}-i\kappa' \cdot x_{2}}= e^{i\frac{\kappa+\kappa'}{2} \cdot
(x_{1}-x_{2}) -i \Delta \cdot \frac{ x_{1}+x_{2}}{2}}.$$ When
diagonalized, and combined with unitarity, imaginary parts of
transition amplitude such as $\Phi$ can also be viewed as parton
probabilities for certain reactions.  Our emphasis retains the
amplitude interpretation.

By superposition the target state is represented as $$|P,s> = \int d
Y \; exp( -iP \cdot Y) |Y,s>,$$ where $Y$ is a center of momentum
($CM$) coordinate.  With a similar step for $|P'>$, we have matrix
elements depending on $e^{-iP \cdot Y+iP' \cdot Y'}=
e^{-i\frac{P+P'}{2} \cdot (Y-Y') +i \Delta \cdot \frac{ Y+Y'}{2} }.$
These steps isolate all dependence on the variable $\Delta$ in \ba &&
\Phi^{\rho \sigma}(\kappa,\kappa')_{P,P'}= \nn \\&& \int dY dY' \, dx_{1}dx_{2}
\bar \Phi e^{-i\frac{P+P'}{2}\cdot (Y-Y')+i\frac{\kappa+\kappa'}{2} \cdot
(x_{1}-x_{2})} e^{-i \Delta \cdot (- \bar Y + \bar b )}, \nn \ea where
$ \bar \Phi_{\rho \sigma}=<Y',s'| \psi_{\rho}(z') \bar
\psi_{\sigma}(z)|Y,s>$ , $\bar Y =(Y+ Y')/2,$ and $$ \bar b=
(x_{2}+x_{1})/2.  $$

The new feature of $GPD$ is given by the $\Delta$
dependence\cite{Soper,Burkardt,buniy}.  From the Fourier expansions, the $\Delta$
dependence is interpreted as a measurement of the conjugate variable
$\bar b-\bar Y$.  This space-time variable is the average {\it spatial
offset of the struck quarks relative to the $CM$ location of the hadrons}, as
resolved on the hard scale $1/\sqrt{Q^{2}}$.  The concept of ``average
spatial offset'' cannot be formulated in the forward limit $\Delta \ra
0$ probed in inclusive experiments.  Thus the object of our study, the
spatial location of the struck quarks inside the proton, cannot in principle
be addressed by the traditional inelastic observables.  Yet the
spatial location is gauge-invariant, local, and underlies all the
sophisticated technical studies of the applicability of
QCD.\cite{RadyWeiss,GPD}.  The dependence on where the hadron
$CM$ is located, $\bar Y$, appears as a trivial overall phase in the
amplitudes; we may therefore set $\bar Y \ra 0$.  Then $\bar b$ is the
average spatial location of the quark correlations relative to that
origin.  This clearly represents a new variable compared to the {\it
separation between the quark fields in correlations} familiar to
workers involved in DIS.

The $\Delta$ dependence, in turn, can be converted back into the
spatial location of the struck quarks by doing the inverse Fourier transform.
Elementary consideration reveals that this amounts to introduction of
a ``focusing lens'' of mathematical sort.  Indeed, the action of an
ordinary (perfect, thin) lens in focusing light consists of Fourier
transforming the {\it ray representation} (momentum) into the {\it
spatial representation}.  The modulus of the spatial amplitude-squared
is the {\it intensity} in the image plane.  Thus, while it is
impossible to focus gamma-rays of GeV energies well with any material
instruments, the focusing can be obtained mathematically, if only the
{\it amplitudes} of the gamma rays emissions are known.

We turn to the underlying reasons that the gamma-ray amplitudes can be
measured.  The cross section $d\sigma \sim \sum_{if } \, M_{if}M_{if}
^{*} $ where $M_{if}$ is the transition matrix element, and $i, \, f $
are indices, such as spins, summed over experimentally, and described
in detail below.  DVCS has $M = M_{BH} +M_{e+target}$ where $M_{BH}$
is the Bethe-Heitler amplitude for the emission from the electrons,
and $M_{e+target}$ is the emission from the target struck by the
electron or positron.  The interference term $d\sigma_{int} =2 Re
\sum_{if } \ [ M_{BH}^{*}(\Delta)M_{e+target}(\Delta) \ ]$ can be
obtained by subtraction since $| M_{BH}|^{2}$ is known.  Symbol
$M_{e+target} $ is already contracted with the amplitude $L^{\nu} $
for $\gamma^{*}$ emission from the lepton beam: with indices explicit,
$M_{e+target}^{\mu} = M_{target;\, s\ , s'; \, \mu \nu}L^{\nu} $ to
emit final state polarization $\mu$.  Symbol $M_{BH} $ includes the
lepton amplitudes contracted with $<p', \, s'| J_{em} ^{\rho} |p, \,
s>$, and depends on $s,\, s'$.  The {\it experimentally measured} form
factors, not an impulse approximation, are used to predict $M_{BH} $. 
Measuring the interference term is practical due to copious electron
radiation, $|M_{BH}| >> |M_{target}| $, which is the generic situation
at moderate energies\cite{HermesCEBAF}.

The interference term can be written \ba \sum_{ss'\, \mu \nu} L^{\nu}
M_{BH; \, ss'}^{\mu *}(\Delta)M_{target;\, s\ , s'; \, \mu
\nu}(\Delta) +c.  \, c.  ; \nn \\ = tr\ [ \, L \, M_{BH}
^{\dagger}(\Delta)\, M_{target}(\Delta) \, \ ] +c.  \, c.  \ea Here $c.  \,
c.  $ indicates the complex conjugate and $tr$ indicates the trace
over the joint indices of proton spins $s,\, s'$ and photon
polarizations $\mu , \, \nu$.  The matrix $\Lambda_{ s \, s'} ^{\mu \nu}
= L^{\nu} M_{BH; \, ss'}^{\mu *} $ can be expanded in an orthonormal
joint basis $m^{j} $ on the space of all matrices, with $tr( (m^{ j}
)^{\dagger} m^{k} ) =\delta^{jk}$.  Choose the first basis element
$m_{ss'; \, \mu \nu}^{1} (\Delta)= \Lambda_{ s \, s'}^{\mu
\nu}/\sqrt{tr(\Lambda^{\dagger} \Lambda)}$ ``parallel'' to the Bethe-Heitler
amplitude itself.  Then $M_{BH}(\Delta)=\mu^{1}(\Delta)m^{1}(\Delta)$
with $\mu^{1}$ a scalar coefficient.  All of the factors being known,
the trace projects out a unique amplitude $ M_{T; \,s\, s'\, }^{ \mu
\nu }(\Delta)= A(\Delta) m_{ss'; \, \mu \nu}^{1} (\Delta)$, where
$A(\Delta) $ is a Lorentz scalar.  We suppressed other dependences to
highlight how the irrelevant $\Delta$-dependence in the scalar
coefficient $\mu_{1}(\Delta) $ has been removed.  Up to normalization
conventions, $M_{T; \,s\, s'\, }^{ \mu \nu }(\Delta)$ equals the
Fourier transform of $<p', \, s'| \, T \ [\, J_{em}^{\mu}(x )
J_{em}^{\nu}(x') \, \ ] \, |p, \, s>$.  The observable amplitude is
the textbook amplitude; the GPD and handbag kernels are constructs
justifying its interpretation.  An unpolarized experiment may extract
only a real or imaginary part of the amplitude, as reviewed below.  A
{\it spin-dependent} amplitude determination is made even from {\it
unpolarized} targets.

This remarkable feature has a physical explanation.  In the emission
of $\gamma(q')$, both the electron beam and the target contribute
amplitudes, which quantum-mechanically interfere.  The virtual photon
strikes the target hard, causing acceleration of the internal
constituents and emission of a real photon {\it from near the struck
point}, in the handbag model.  Simultaneously the electron beam acts
as a known coherent reference sources, and emits a photon with known
quantum numbers.  Amplitudes orthogonal to the reference source may be
emitted by the target, but they cannot be observed by interference:
the experimental conditions of the lepton beam and polarization sums
{\it select} the survivor among all possible amplitudes.

The interference amplitude is odd in the sign of the lepton charge,
while $M_{BH}$ has no absorptive part, so that the real part of
$M_{target}$ is immediately extracted by electron-positron charge
asymmetry.\cite{BrodskyCloseGunion,DGPR}.  Angular momentum analysis
shows that the imaginary part of $M_{target}$ can be obtained
\cite{DGPR} by flipping longitudinal polarization of the lepton.  The
spin-dependent amplitudes obtained from unpolarized targets will be
incomplete, and not the {\it most general} amplitudes, which await
polarized targets.  Moreover it has been shown\cite{bel} that
approximate information on the $\Delta -$ dependence of the function
$\Phi$ can be obtained without a full set of polarization
measurements.  Whatever information at the amplitude level that can be
obtained immediately leads to a corresponding {\it image} of the
target under the conditions of the experiment, which should be
immensely informative.

Step by step, to implement the procedure one needs to:

\begin{itemize}

\item $\bullet$ Extract amplitudes for the reaction and fit them with
smooth functions of $\Delta$.  Not all linearly independent amplitudes
are needed: Spin averaging the target, for example, mixes independent
images that might have otherwise been separated.  Spin-dependence,
using a polarized target, remains extremely interesting and is also
technically feasible.  Amplitudes can be extracted for fixed values of
the skewness parameter, or they may be integrated over a region of
skewness: longitudinal information is intrinsically smeared over a
scale comparable to the target thickness.  We must warn that skewness
values close to zero or one will involve different physics, as for
$x_{Bj}$ dependence in DIS. Each procedure generates a different photo
from the $\DD_{T}$ dependence of what was observable under the
observing conditions.

\item $\bullet$ The amplitudes should be measured in $t$-bins from
$-t_{min}$ to $-t \sim 1 \, GeV^{2}$.  Fit the amplitudes in $\DD_{T}$
(or $t$ if symmetry exists) and generate the Fourier transform in
$\vec b_{T} $ to get a profile amplitude function $f(\vec b_T)$.  The
interpretation of $f(\vec b_T)$ is the amplitude to "find" quarks at
$\vec b_{T}$ in an image plane after focusing by an idealized lens.
The term ``find'' means to extract and reinsert quark fields in
correlation, or do the same with a quark-antiquark pair, depending on
longitudinal momenta, as resolved in convolution with the handbag
kernel.

\item $\bullet$ Square the profile amplitude, producing $|f(\vec
b_T)|^{2},$ which is positive, real-valued, and corresponds to the
``image'', a weighted probability to find quarks in the transverse
image plane.  Such probabilities, like conventional parton
probabilities, represent the features of universal matrix elements;
when measured the same way between different reactions, they must be
the same picture.  When measured the same way at different $Q^{2}$,
they are the same picture viewed at different spatial resolutions.
This suggests a new era of systematic comparison of partonic structure
of hadrons, greatly more detailed than that of the past two decades.

\item $\bullet$ The deuteron\cite{deut} is a superb example where the
strategy can be applied.  Its wave function to locate the proton and
neutron are known, but its wave function to locate {\it quarks and
gluons} has never been measured.  Larger nuclei are just as amenable
to the process: one should arrange for events to be quasi-elastic on
the nucleons, below the threshold for pion emission.

\end{itemize}

The images so obtained are transmission photographs, which brings up
certain details of the longitudinal coordinate.  Recall that
$\Delta^+$ is comparable to the largest scales in the problem, namely
$P^+$ and $\sqrt{Q^2}$.  The conjugate spatial coordinate locating the
struck quarks, $\bar b^- \sim b_0 - v b_3 \sim 1/\Delta^+$, where $v
\sim c$ is the lon  gitudinal velocity of the quark, is isolated to no better than
the spatial resolution.  Consistently, the proton is optically thin, a
relativistic pancake in the center of mass frame consisting of about a
single longitudinal layer of quarks, antiquarks and gluons.  One may
retain the $\Delta^+$ information and do a form of {\it tomography}
through the longitudinal slices of the proton, granting the
limitations of resolution, or one may integrate over $\Delta^+$. 
Various coordinates are possible: Soper's ``center of $p^+$'' variable
may be useful\cite{Soper}.

Bjorken scaling in $Q^{2}$ is assumed: scaling violations are loss of
resolution.  For our purposes logarithmic scaling violations are ``old
physics'' .  In that event integrating over $Q^{2}>Q_{min}^{2}\sim
GeV^{2}$ will be useful in accumulating statistics.

The measurement also depends on the probe.  The handbag kernel
\cite{GPD} $1/(x-\zeta/2+ i \epsilon)$ depends on the longitudinal
momentum fraction $x$ via convolution with the amplitude to find
the quarks.  By the convolution theorem, the effect is the {\it
product} of the amplitude and the kernel in the conjugate $ z-v t$
spatial variable: with the kernel then a Heaviside step function.  The
step function is the action of an instantaneously opening camera
``shutter'', which naturally affects the photographic resolution in
the longitudinal direction.  The longitudinal kernel differs for
gluons, which couple through quark pairs and dominate at small $x$,
while the rest of the interpretation holds.  Concerns about spectator
interactions\cite{BHS} do not disturb the extraction of an image,
which represents what was observable under the conditions of the
experiment.

Remarkably, the transverse coordinates are absolutely decoupled.  The
natural image plane is the transverse spatial one cutting across the
face of the proton and resolved on the scale of $1/\sqrt{Q ^2}$.
Current technology allows $1/\sqrt{Q ^2}<< \,Fm $ and therefore {\it
femto-photography of the interior structure of the proton.} The data
to evaluate these photos already exists, or will exist soon.  What
will the images look like?  Surprisingly, the answer is unknown.

Despite Hofstadter's beautiful measurements, and their
continuation\cite{formfactors} ``\ldots achieved by improving unrelentingly
(the) methods and equipment in the course of time'' \cite{IWwaller},
the proton has relativistic constituents, and the non-relativistic
impulse approximation is no longer credible at the 100 MeV scale of
form factor structure.  The proton's size inferred for 40 years has
been intermixed with the photon's spatial resolution, and also lacks
independent verification.  Consistency of other measures, such as
those of atomic physics, are circular and add no information.  A host
of strong-interaction measures still lack the precision to be
definitive.  Surely, the proton's image cannot differ too much from
the small round dot of about 1 Fm in diameter, so long imagined.  But
is the dot exactly 1 Fm, 2 Fm, or 1/2 Fm in size?  Is the dot round?
Given the fact that the transverse spin of the proton breaks
rotational symmetry, the proton's image need not be circular: quark
orbital angular momentum, an exceptionally controversial topic at
present, can reveal itself in breaking of rotational
symmetry.\cite{buniy}.  Are the up quarks, which have charge
2/3, located on the outside, inside, or with uniform distribution in
the proton?  Nobody knows.  Where are the strange quarks, the gluons,
and how is the spin distributed among all these species?  We find that
DVCS can resolve such questions, because a new principle using the
momentum transfer, and not the detected photon wavelength, fixes the
spatial resolution.

If current technology can image a proton, what are the limits of this
new technique?  The proton's size is about $ 10^{-7}$ of the size of
interesting biological molecules, which billion dollar facilities
explore via Bragg diffraction.  Yet it is inconceivable to scatter
from a biological molecule {\it elastically} with a high energy
photon.  Assuming nuclear locations alone are sought, one needs $|\vec
\Delta| < \sim \hbar/A^{o}\sim keV$; with relativistic beams in the 10
MeV-GeV range, a relative precision of order $10^{-5}-10^{-7}$: and we
believe, beyond current technology.  Nevertheless the photography of
microscopic objects has never before been possible with the wavelength
of the photon separate and independent from the resolution, so
unforeseen technological applications may exist.

\begin{figure}

\epsfxsize=3in\epsfysize=1.5in

\epsfbox{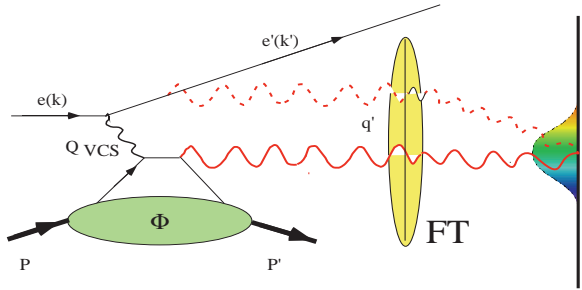}

\caption{ Diagrammatic representation of the ``handbag'' diagram
probed in DVCS; the crossed photon case is omitted.  Emission of the
final state photon from the target (solid $q'$ wave) and lepton beam
(dashed $q'$ wave) with coherent interference is also shown.  A
Fourier transform (``FT'') represented by a ``lens'' creates a
probability map of the struck quark probability (shaded) on the
transverse spatial image plane.}

\end{figure}

\begin{figure}

\epsfxsize=3in\epsfysize=1.2in

\epsfbox{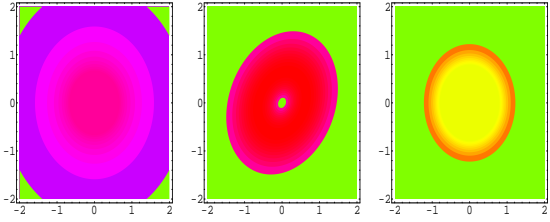}

\caption{Typical photos, or model images made from transverse position space
($\vec b_{T}$) amplitude squared; scale in Fermi.  Images made on
log-color plots.  Left to right: Hofstadter's rendition; a model with
transverse spin, plus a quadratic hole from 3-quark correlations; a
proton one-half the size and more concentrated than traditional.}

\end{figure}

\medskip

{\it Acknowledgements} Work supported in part under Department of
Energy Grant.  CphT, Ecole Polytechnique is UMR 7644 of CNRS. We
acknowledge useful discussions with P. Jain, G. van der Steenhoven, P.
Mulders, M. Gar$\c $con, M. Diehl, N. D'Hose, P. Kroll, R. Jacob and P. A. M.
Guichon.

\end{document}